\documentclass[letter, traditabstract]{aa} 
\usepackage{graphicx}
\usepackage{natbib}
 \usepackage{longtable,lscape}

\newcommand{\kms}{{\rm km}\,{\rm s}^{-1}}
\newcommand{\msun}{{\rm M}_\odot}

\newcommand{\vrotsigma}{V_{\rm rot} \sin{i} \ /\sigma_{\rm 1D}}

\def\apj{ApJ}
\def\mnras{MNRAS}

\newcommand{\mvphi}{\langle v_\phi\rangle}
\newcommand{\vrms}{v_{\rm rms}}
\newcommand{\vrotsini}{V_{\rm rot}\sin\,i}
\newcommand{\vrotsigoned}{V_{\rm rot}/\sigma_{\rm 1D}}
\newcommand{\sigoned}{\sigma_{\rm 1D}}

\newcommand{\sini}{\sin{i}}

\usepackage{txfonts}
\begin{document}

  \title{The VLT-FLAMES Tarantula Survey\thanks{Based on observations collected at the European Southern Observatory under program ID 182.D-0222}}

   \subtitle{VI. Evidence for rotation of the young massive cluster R136}

   \author{V. H\'{e}nault-Brunet\inst{1} \and M. Gieles\inst{2}  \and C. J. Evans\inst{3,1} \and H. Sana\inst{4}  \and N. Bastian\inst{5} \and J. Ma{\'\i}z Apell{\'a}niz \inst{6}  \and W. D. Taylor\inst{1} \and N. Markova\inst{7} \and E. Bressert\inst{8,9,10} \and A. de Koter\inst{4} \and J. Th. van Loon\inst{11} }

   \institute{Scottish Universities Physics Alliance (SUPA), Institute for Astronomy, University of Edinburgh, Blackford Hill, Edinburgh, EH9 3HJ, UK ; \email{vhb@roe.ac.uk}              
              \and
           Institute of Astronomy, University of Cambridge, Madingley Road, Cambridge, CB3 0HA, UK
              \and
              UK Astronomy Technology Centre, 
           Royal Observatory Edinburgh, 
           Blackford Hill, Edinburgh, EH9 3HJ, UK
           \and
           Astronomical Institute `Anton Pannekoek', University of Amsterdam, Postbus 94249, 1090 GE, Amsterdam, The Netherlands
           \and
           Excellence Cluster Universe, Technische Universit\"{a}t M\"{u}nchen, Boltzmannstr. 2, D-85748 Garching, Germany
           \and
   Instituto de Astrof{\'\i}sica de Andaluc{\'\i}a-CSIC, Glorieta de la Astronom\'{\i}a s/n, E-18008 Granada, Spain
  	\and
	           Institute of Astronomy with NAO, Bulgarian Academy of Sciences, PO Box 136, 4700 Smoljan, Bulgaria
	\and
	School of Physics, University of Exeter, Stocker Road, Exeter EX4 4QL, UK
	\and
	European Southern Observatory, Karl-Schwarzschild-Strasse 2, D87548, Garching bei M\"{u}nchen, Germany
	\and
	Harvard-Smithsonian CfA, 60 Garden Street, Cambridge, MA 02138, USA
	\and
	Astrophysics Group, Lennard-Jones Laboratories, Keele University, Staffordshire ST5 5BG, UK
                 }

   \date{Received ; accepted }

 
   \abstract
   {Although it has important ramifications for both the formation of star clusters and their subsequent dynamical evolution, rotation remains a largely unexplored characteristic of young star clusters (few Myr). Using multi-epoch spectroscopic data of the inner regions of 30~Doradus in the Large Magellanic Cloud (LMC) obtained as part of the VLT-FLAMES Tarantula Survey, we search for rotation of the young massive cluster R136. From the radial velocities of 36 apparently single O-type stars within a projected radius of 10\,pc from the centre of the cluster, we find evidence, at the 95\% confidence level, for rotation of the cluster as a whole. We use a maximum likelihood method to fit simple rotation curves to our data and find a typical rotational velocity of $\sim$3\,$\kms$. When compared to the low velocity dispersion of R136, our result suggests that star clusters may form with at least $\sim$20\% of the kinetic energy in rotation.}


\authorrunning{V. H\'{e}nault-Brunet et al.}
\titlerunning{VFTS VI. Evidence for rotation of R136}

   \maketitle
%

\section{Introduction}

Despite their spherical shape, Milky Way globular clusters (GCs) rotate with amplitudes up to half the 1D velocity dispersion \citep[$0\lesssim\vrotsigma\lesssim0.5$, e.g.][]{1997A&ARv...8....1M}, so their amount of rotational energy is typically not dominant but also not negligible. Numerical studies have shown that this rotation has an important influence on star clusters by accelerating their dynamical evolution, for example by speeding up the collapse of the core through the gravogyro instability or by significantly increasing the escape rate for clusters in a strong tidal field \citep[e.g.][]{1999MNRAS.302...81E, 2002MNRAS.334..310K, ernst2007}.

Most of the rotation signatures  are found through radial velocity (RV) studies, but rotation has also been confirmed in the plane of the sky  for  $\omega$ Centauri and 47 Tucanae \citep[][respectively]{2000A&A...360..472V, 2010ApJ...710.1032A}. RV studies are now able to measure rotational amplitudes in GCs below 1 $\kms$ and despite these precise measurements rotation has not been detected in some clusters \citep[e.g.][]{2010MNRAS.406.2732L}, although this could also be an inclination effect. 

It is unclear what the origin of the rotation is in some of these old clusters. It could be the result of the merging of two clusters \citep{2003ApJ...589L..25B} or imprinted during the formation process. Observational input is now getting sufficiently abundant to look for correlations between rotational amplitude and other cluster properties. \citet{2012A&A...538A..18B} report a correlation between horizontal branch (HB) morphology and $\vrotsigma$ in a sample of 20 globular clusters. Given that metallicity is the first parameter determining HB morphology, this in turn suggests a correlation between $\vrotsigma$ and metallicity, such that clusters with higher metallicity have greater fractions of energy in rotation. Since a higher metallicity in a gas implies a higher efficiency of energy dissipation by atomic transitions, this then hints at a significant role of dissipation in the process of cluster formation \citep[e.g.][]{2010ApJ...724L..99B}. Rotation may therefore be intimately linked to the formation of clusters.

Little is known about rotation in young clusters, partially because it is very challenging to measure accurately $\sigma_{\rm 1D}$ given the high multiplicity fraction of massive stars \citep[e.g.][]{vhb2012a}. \citet{1982MNRAS.199..565F} found an age-ellipticity relation for clusters in the Large Magellanic Cloud (LMC) with older clusters presenting less elongated shapes, which was interpreted as internal evolution erasing any asymmetry stemming from the violent relaxation process of the formation \citep[although see][]{goodwin1997}. However, rotation and ellipticity are not necessarily equivalent. Ellipticity can be due to rotation \citep[e.g. $\omega$ Cen;][]{meylan1986} but also to velocity anisotropy \citep{stephens2006, Henon1973}, and rotating clusters can be spherical \citep{LB60, meza2002}.

Marginal evidence for rotation was found for the young (few 100 Myrs) Galactic cluster GLIMPSE-C01 with an amplitude of $\vrotsigma\simeq0.2$ \citep{2011MNRAS.411.1386D}.  A rotational signal in the RVs was also detected in the $\sim$100 Myr cluster NGC\,1866 \citep{fischer1992} and in the $\sim$50\,Myr binary cluster NGC\,1850 \citep{fischer1993}, both in the LMC. To really confirm whether clusters form with a significant amount of angular momentum, we need to look for an even younger cluster. The young massive cluster (YMC) R136 in the 30 Doradus star forming region in the LMC is an ideal target to establish this.  With an estimated mass of about $10^5\,\msun$ \citep{2009ApJ...707.1347A} and its sub solar metallicity, it may at some stage resemble a typical metal-rich GC as we find them in the Milky Way Bulge. With an age of less than 2\,Myr \citep{koterheap, massey1998, Crowther2010}, it is so young that any rotation needs to be attributed to the formation process, be it from merging of sub-clusters or directly from the angular momentum of the progenitor cloud. A rough estimate of the half-mass relaxation time ($t_{\rm rh}$) of R136 can be obtained by assuming $N=10^5$ stars and a half-mass radius of 2.27\,pc, which is found from multiplying the half-light radius of 1.7\,pc \citep[e.g.][]{vhb2012a} by 4/3 \citep{spitzer:1987}. Following the formula of \citet{spitzerhart1971}, we obtain $t_{\rm rh}\simeq366$\,Myr, so relaxation would not have had time to erase the original signature of rotation.

Here we report on evidence for rotation of R136 deduced from RV measurements of massive stars obtained as part of the VLT-FLAMES Tarantula Survey \citep[VFTS;][]{evans:2011}. We briefly present the data in Sect.~\ref{data} and describe our analysis of the rotational signature in Sect.~\ref{analysis}. We discuss the implications of the rotation of R136 for cluster evolution in Sect.~\ref{discuss}, and present our conclusions in Sect.~\ref{conc}.

   \begin{figure}[!h]
   \centering
   \includegraphics[width=8.5cm]{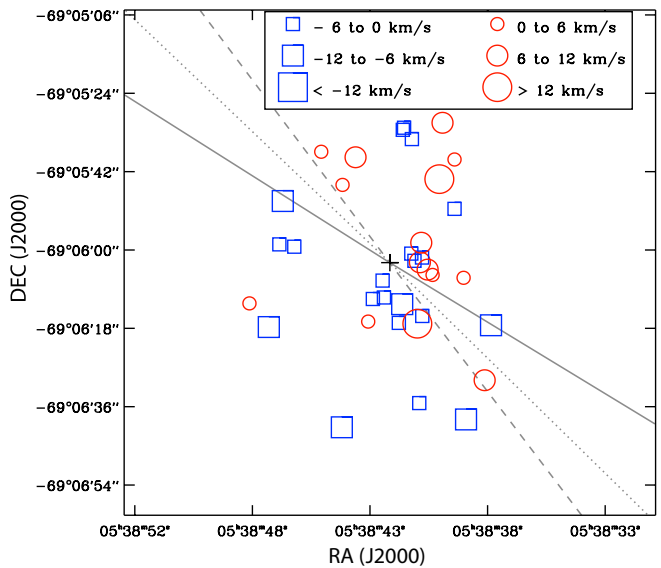}
      \caption{Illustration of the positions and RVs of the stars considered in this study. Symbol sizes denote the magnitude of the stellar velocities with respect to the average cluster velocity. The solid, dotted, and dashed lines correspond to the optimal rotation axis determined for models with a constant rotational velocity, a constant rotation rate, and a more realistic rotation curve (see Sect. \ref{analysis}), respectively.}
         \label{rot_circles}
   \end{figure}

\section{Data \label{data}}

The data used in this work consist of RV measurements and their uncertainties for 36 apparently single O-type stars within a projected radius of 10\,pc from the centre of R136 (adopted here as the position of the star R136-a1: $\alpha$ = 5$^{\rm h}$38$^{\rm m}$42$^{\rm s}$. 39, $\delta$ = $-$69$\degr$06'02.''91, J2000). These are based on observations from the VFTS, for which at least five epochs were obtained for five different pointings of the FLAMES--ARGUS integral-field unit in the central arcminute of 30 Dor, in addition to nine configurations of the Medusa fibre-feed to the Giraffe spectrograph in the surrounding Tarantula Nebula over a 25 arcmin diameter field-of-view \citep{evans:2011}. The RV and variability analysis, based on Gaussian fitting of selected helium stellar absorption lines, is presented in \citet{vhb2012a} for the ARGUS data and in \citet{sana_vfts} for the Medusa data. We retain here only the RVs of stars showing no significant variability throughout all epochs, and we apply a 3\,$\sigma$ clipping centered on the mean velocity of the cluster, with $\sigma=6$\,$\kms$, the observed line-of-sight velocity dispersion of the cluster \citep[i.e. before correction of the velocity dispersion for undetected binaries, see][]{vhb2012a}. This yields a total of 16 ARGUS sources, all within 5\,pc from the centre, and 20 Medusa sources, all between 5 and 10\,pc except two between 4 and 5\,pc. The identification numbers, coordinates, spectral types, and RVs of these stars are listed in Table 3 of \citet{vhb2012a}. In Fig.~\ref{rot_circles}, we schematically present their positions and distribution of RVs with respect to the mean RV of the cluster.

\section{Analysis \label{analysis}}

To look for the signature of rotation, we fit rotating models to our set of measured RVs by maximizing the logarithm of the likelihood function \citep{bevington}

\begin{equation}
M = \ln {\mathcal L} = \ln \left[ \prod_{i=1}^{N} P_i \right] = \sum_{i=1}^{N} \ln P_i.
\end{equation}

\noindent Given our relatively small data set, we only consider three simple models: constant rotational velocity amplitude, constant rotation rate, and finally a more realistic model with solid body rotation in the inner parts peaking near 2 half-mass radii followed by a decline. For these models, the probability density function $P_i$ of a measurement RV$_{i}$ with uncertainty $\sigma_i$ given a position angle PA$_0$ for the rotation axis and a line-of-sight velocity dispersion $\sigma_{\rm 1D}$ can be written as

\begin{equation}
P_i = \frac{1}{ \sqrt{2\ \pi} \sqrt{\sigma_i^2 + \sigma_{\rm 1D}^2} } \exp{\left[ - \frac{1}{2} \left(\frac{{\rm RV}_i - V_{\rm rot} \sin{i}}{\sqrt{\sigma_i^2 + \sigma_{\rm 1D}^2}}\right)^2 \right]}, 
\label{P_dist}
\end{equation}

\noindent{where $V_{\rm rot} \sin{i}$ is a constant for a model with fixed rotational velocity. For a model with constant rotation rate, $V_{\rm rot} \sin{i}$ depends on the rotation rate ($\Omega$) and the distance from the rotation axis ($X_j$) such that $[V_{{\rm rot}} \sin{i}]_{j}=\Omega \ X_j$. For the physically motivated model, we adopt a function of the form $[V_{{\rm rot}} \sin{i}]_{j}= A/2 \cdot  X_j / (1 + (X_j/4)^2)$, where $X_j$ is the distance to the rotation axis in pc and $A$ is the maximum rotational velocity which we assume is at 4\,pc for simplicity (the half-mass radius of R136 is about 2\,pc). This rotation curve captures the general behaviour seen in simulations of rotating clusters \citep[e.g.][]{2002MNRAS.334..310K}. Note that because the velocity dispersion profile of R136 is relatively flat \citep{vhb2012a}, this function also describes $V_{\rm rot}/\sigma$. We define the position angle (with respect to the centre of the cluster) as increasing anti-clockwise in the plane of the sky from North (PA$=0^{\circ}$) towards East (PA$=90^{\circ}$). We adopt negative rotational velocities for position angles between PA$_0$ and PA$_0$+180$^{\circ}$. For simplicity, we assume that $\sigma_{\rm 1D}$ is constant across the radius range considered. This $\sigma_{\rm 1D}$ is largely due to cluster dynamics, but contains a small contribution from the orbital motions of undetected binaries \citep[1-2 $\kms$;][]{vhb2012a}, effectively adding noise to the rotation signature we are trying to detect.

   \begin{figure*}[!t]
   \centering
\includegraphics[width=17.5cm]{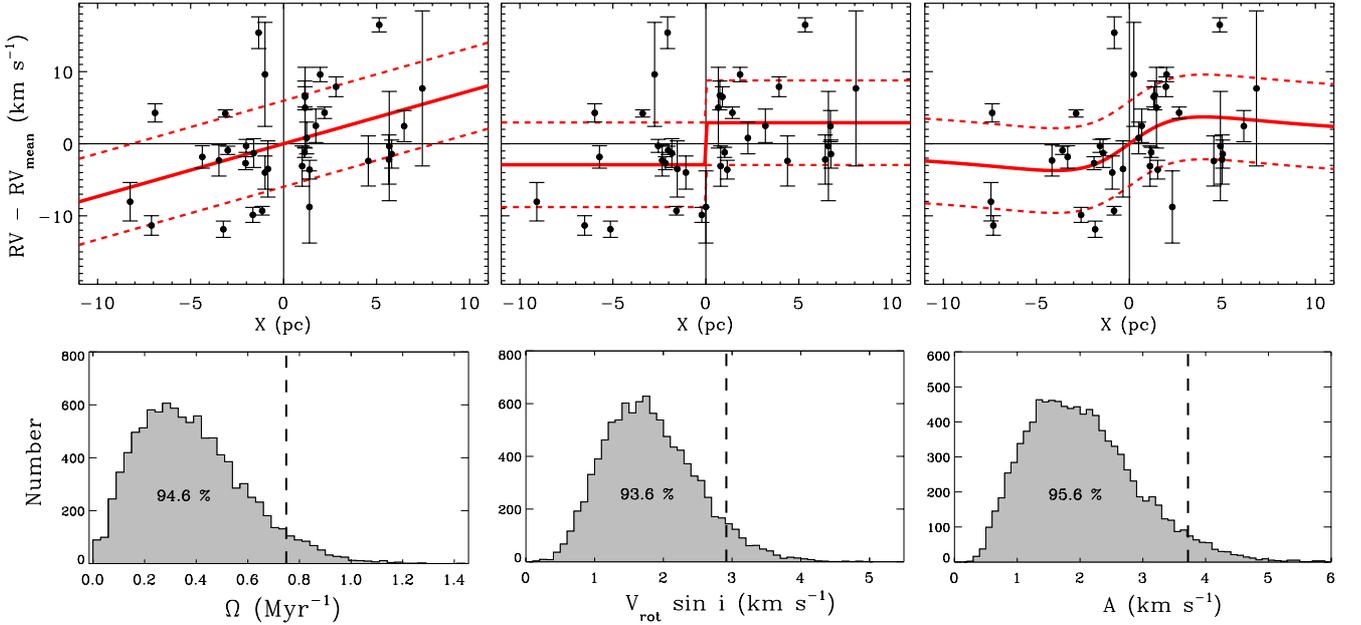}

      \caption{{\it Top:} RV in the system of the cluster as a function of distance $X$ from the optimal rotation axis for models with constant rotation rate (left), constant rotational velocity (centre), and the more realistic rotation curve discussed above (right). The best-fit rotation curves are shown as solid lines, and $\pm \sigma_{\rm 1D}$ envelopes are represented by dashed lines. {\it Bottom:} Histogram of amplitudes $\Omega$ (left), $V_{\rm rot}$ (centre) and $A$ (right) from 10\,000 Monte Carlo simulations of RV distributions with no rotation. The optimal values for the observed RV configuration are indicated by dashed vertical lines. Confidence levels of 94.6\%, 93.6\% and 95.6\% are found respectively for the best-fit amplitude of each model.}
         \label{rot_curve}
   \end{figure*}

The values of the parameters that maximize the likelihood function are PA$_0$=57$\pm$15$^{\circ}$, $V_{\rm rot} \sin{i}$=2.9$\pm$1.0\,km\,s$^{-1}$ and $\sigma_{\rm 1D}$=5.9$\pm$0.8\,km\,s$^{-1}$ for the constant rotational velocity model, PA$_0$=44$\pm$18$^{\circ}$, $\Omega$=0.75$\pm$0.22\,Myr$^{-1}$ and $\sigma_{\rm 1D}$=6.0$\pm$1.1\,km\,s$^{-1}$ for the constant rotation rate model, and PA$_0$=36$\pm$14$^{\circ}$, $A$=3.7$\pm$1.3\,km\,s$^{-1}$ and $\sigma_{\rm 1D}$=5.9$\pm$0.8\,km\,s$^{-1}$ for the final model. The best-fit rotation curves are shown in Fig.\,\ref{rot_curve}. The uncertainties were estimated using Monte Carlo calculations on simulated data sets comparable to our measured data (i.e. with the same rotational signal). We also obtained consistent uncertainties by considering the parameter change necessary to decrease $M$ by $\Delta M= \frac{1}{2}$ from its value at the maximum \citep{bevington}. The maximum value of $M$ is slightly higher for the constant rotational velocity model compared to the two other models, but likelihood ratio tests showed that the difference is not significant. Thus, we currently cannot favour one model over the others.

To establish the significance of the detected rotational signal, we performed Monte Carlo simulations and applied the maximum likelihood method above to 10\,000 random distributions of velocities (i.e. non-rotating systems). The adopted size and spatial distribution of the simulated populations were taken to be the same as the observed sample in order to be sensitive to possible biases introduced by our non-uniform spatial sampling. The velocities were drawn from a Gaussian distribution with $\sigma$ determined by the observed line-of-sight velocity dispersion. For each simulated star, observational noise was added based on the RV uncertainty of the observed star at the same location, again to take into account the biases introduced by our non-uniform data set. The distribution of $\Omega$, $V_{\rm rot} \sin{i}$ and $A$ from 10\,000 such simulations is shown in Fig.\,\ref{rot_curve}. The distributions do not peak at zero because the limited number of stars and the measurement noise generally result in a non-zero amplitude. Our best-fit values are located in the right tail of these distributions and the corresponding confidence level of the detection is around 95\% for all three models.

The analysis outlined above was also performed on a subsample from which supergiant candidates and stars with possible composite spectra were excluded as their RVs might be inaccurate \citep[see][]{vhb2012a}. Very similar results were obtained, with best-fit parameters all within 4\% of those previously determined and a confidence level about 1\% lower because of the smaller number of measurements. As an additional check, we ran the analysis on different subsamples of apparently single O-type stars from VFTS using the measurements of \citet{sana_vfts}, including stars farther away from R136 than in our main sample. The analysis of these subsamples suggests that the rotation signal is dominated by the stars in the inner regions and does not result from a velocity gradient across the field on a larger scale in the surrounding OB association. For example, when considering the sample of stars between 10 and 20\,pc from the centre (37 stars), we find that $V_{\rm rot} \sin{i}$ goes down to 1.8\,$\kms$ for a constant rotational velocity model and the confidence level of the rotational signature is only 51\%. Note that stars that are part of the surrounding 30 Doradus region and not members of R136 could contaminate our sample. If we assume that the outer component of the double-component EFF fit \citep{1987ApJ...323...54E} to the light profile of R136 \citep{2003MNRAS.338...85M} is due to the surrounding association, then we may expect these stars to contribute to $>50\%$ of the sample beyond 5 pc from the centre and to dilute the rotational signal in the outer regions of the cluster.

We also looked for the signature of rotation with a method commonly used in studies of globular clusters. This method consists of dividing the sample with a line passing through the centre of the cluster and computing the difference in the mean RV between the two subsamples of stars as a function of the position angle of the dividing line \citep[e.g.][]{cote1995, 2012A&A...538A..18B}. Results very similar to those reported above were obtained for the position of the rotation axis and the rotational amplitude, but the maximum likelihood method has the advantage of directly comparing the data to models without binning.

\section{Discussion \label{discuss}}

Given that $\sigoned\simeq5\pm1\ \kms$ for R136 after correcting for undetected binaries \citep{vhb2012a} and that the mean rotational velocity in the radius range considered is $\simeq3\pm1$\,$\kms$ for the three best-fit rotation curves shown above, our analysis implies that $\vrotsigma\simeq0.6\pm0.3$, which is somewhat larger than what is typically observed in globular clusters \citep[e.g.][]{1997A&ARv...8....1M}. We have to keep in mind that $\vrotsini$ is itself a projected quantity, and so is $\sigoned$. We would like to know how these quantities relate to the 3D quantities $\mvphi$ and $\vrms$. Here $\mvphi$ is the mean tangential velocity component in the $\phi$ direction, used as the rotational component of the velocity vector in spherical coordinates for a cluster rotating about the $z$ axis, and $\vrms$ is the root-mean square of the velocities.

We can estimate that $\vrotsigoned$ is typically about $10\%$ larger than $\mvphi/\vrms$. This can be understood as follows. For a circular orbit with unit velocity and the line of sight along the orbital plane ($\sini=1$) the mean RV component is $2/\pi$. For an isotropic velocity dispersion the observed component is $\sigoned=\vrms/\sqrt{3}$. This means that $[\vrotsigoned]/[\mvphi/\vrms]\simeq 2\sqrt{3}/\pi\simeq1.1$. If $\sini=1$ then we would have $\mvphi/\vrms \simeq0.9\vrotsigoned$ and a lower limit on $\mvphi/\vrms$ of 0.5$\pm$0.2. A ratio of $\mvphi/\vrms=0.5$ implies a ratio of kinetic energy in rotation over kinetic energy in random motions of 0.25, which in turn implies that at least 20\% of the total kinetic energy is in rotation. If instead we assume that $\sini=2/\pi$ (the average of $\sini$ assuming a random distribution of inclination angles), we have $\mvphi/\vrms=0.9$ and 45\% of the total kinetic energy in rotation. 
The approximate criterion for stability against non-axisymmetric perturbations provided by \citet{OP1973} states that the ratio of the rotational kinetic energy over the potential energy should not exceed $| T_{\rm rot}/W | <0.14$. If we assume virial equilibrium, then $T = -0.5 W$, so for $T_{\rm rot}=0.20 \ T$ and $T_{\rm rot}=0.45 \ T$ this implies $| T_{\rm rot}/W | =0.1$ and 0.225, respectively. Although not a rigorous test, this suggests that a low inclination is perhaps more reasonable.

If the cluster was flattened by rotation (although recall our previous words of caution about ellipticity and rotation), we would expect to see peaks in the azimuthal density profile, with density minima coinciding with the rotation axis (i.e. PA$_0$$\sim$45$^{\circ}$ and 225$^{\circ}$) and density maxima 90$^{\circ}$ away from the rotation axis (i.e. PA$_0$$\sim$135 and 315$^{\circ}$). However, only one minimum is seen at a position angle of $\sim$100$-$120$^{\circ}$ in the $K_s$- and $H$-band azimuthal density profiles of R136 by \citet{2010MNRAS.405..421C}.

Given the very young age of R136, the cause of the rotation needs to be looked for in the details of the formation process of the cluster. This is a short and complicated phase in which various physical processes, with their respective  time-scales, operate simultaneously.
A logical starting point is to see whether giant molecular clouds (GMCs), the birth sites of YMCs, rotate. 
 \citet{2003ApJ...599..258R} found that GMCs in M33 have non-zero angular momentum. From simple arguments based on differential rotation in a galactic potential and self gravity we expect that the rotation of GMCs should be prograde with the orbit in the galaxy. However,  \citet{2003ApJ...599..258R}
found that 40\% of the GMCs have retrograde motions, which supports a scenario in which GMCs form through both agglomeration and self gravity  and the angular momentum is the result of the clumpiness of the gas \citep{2011MNRAS.417.1318D}. As we pointed out above, a merger is another way to give rise to rotation. Interestingly, \citet{sabbi} found a dual structure in the density of low mass stars in R136 that possibly hints at a relatively recent merger event of the main core of R136 and a second clump or cluster, but only about three of our targets are located in this second clump.

To establish whether the rotation in old GCs is a remnant of their formation, we need to know if the 
angular momentum can survive for a Hubble time of dynamical evolution.
 During the evolution, angular momentum is diffused outward  \citep{1985IAUS..113..285F, 1999MNRAS.302...81E} and ultimately lost through the escape of stars with high angular momentum \citep{1958SvA.....2...22A,1976ApJ...210..757S}. This process operates on a relaxation time and the angular momentum reduces after a fixed number of elapsed relaxation times. The relaxation time of expanding clusters grows roughly linearly in time, which makes the number of elapsed relaxation times grow slowly, namely as a logarithm of the age. Because the majority of GCs are in this expansion phase \citep{2011MNRAS.413.2509G}, we may expect rotational signatures to still be present after a Hubble time because they have not evolved enough. For clusters that have entered the `mass-loss' phase, we do not expect the rotation to survive.

\section{Conclusion \label{conc}}

We presented evidence that the young massive cluster R136 is rotating with a rotational velocity amplitude of about 3\,km\,s$^{-1}$, which implies that at least $\sim$20\% of its total kinetic energy is in rotation. Obviously, RV measurements of more stars in this cluster would be desirable to better populate the rotation curve and confirm the rotational signal with a confidence level higher than the current 95\%. Given the young age of R136, our results suggests that star clusters may form with a significant amount of angular momentum. This will place useful constraints on models of cluster formation. We finally argued that the rotation of globular clusters could originate from their formation, but this is clearly a topic where more detailed numerical investigations are welcome.

\begin{acknowledgements}
We would like to thank the referee for constructive feedback. We also wish to thank Sergey Koposov for useful discussions. VHB acknowledges support from the Scottish Universities Physics Alliance (SUPA) and from the Natural Science and Engineering Research Council of Canada (NSERC). MG acknowledges financial support from the Royal Society. NB was supported by the DFG cluster of excellence `Origin and Structure of the Universe' (www.universe-cluster.de). JMA acknowledges support from [a] the Spanish Government Ministerio de Educaci\'on y Ciencia through grants AYA2010-15081 and AYA2010-17631 and [b] the Consejer{\'\i}a de Educaci{\'o}n of the Junta de Andaluc{\'\i}a through grant P08-TIC-4075. NM was supported by the Bulgarian NSF (DO 02-85). \end{acknowledgements}

\end{document}